\documentclass{article}

\usepackage{arxiv}

\usepackage[utf8]{inputenc} 
\usepackage[T1]{fontenc}    
\usepackage{hyperref}       
\usepackage{url}            
\usepackage{booktabs}       
\usepackage{amsfonts}       
\usepackage{nicefrac}       
\usepackage{microtype}      
\usepackage{graphicx}
\usepackage{cite}
\usepackage{doi}

\usepackage{appendix}
\usepackage{amsmath}

\begin{document}
\title{Outlier Detection in Large Radiological Datasets using UMAP}
%
%


\author{ Mohammad Tariqul Islam and Jason W. Fleischer\thanks{Corresponding Author} \\
	Department of Electrical and Computer Engineering\\
	Princeton University\\
	Princeton, NJ 08544, USA \\
	\texttt{\{mtislam,jasonf\}@princeton.edu} 
}

\maketitle              
\begin{abstract}
The success of machine learning algorithms heavily relies on the quality of samples and the accuracy of their corresponding labels. However, building and maintaining large, high-quality datasets is an enormous task. This is especially true for biomedical data and for meta-sets that are compiled from smaller ones, as variations in image quality, labeling, reports, and archiving can lead to errors, inconsistencies, and repeated samples. Here, we show that the uniform manifold approximation and projection (UMAP) algorithm can find these anomalies essentially by forming independent clusters that are distinct from the main (“good”) data but similar to other points with the same error type. As a representative example, we apply UMAP to discover outliers in the publicly available ChestX-ray14, CheXpert, and MURA datasets. While the results are archival and retrospective and focus on radiological images, the graph-based methods work for any data type and will prove equally beneficial for curation at the time of dataset creation. 
\end{abstract}
\keywords{x-ray \and data visualization \and data curation \and neighbor embedding.}
\section{Introduction}
\label{sec:intro}
\let\thefootnote\relax\footnotetext{Accepted in MICCAI-2024 Workshop on Topology- and Graph-Informed Imaging Informatics (TGI3)}
A prominent reason behind the current success of machine learning-based disease detection is the availability of large medical datasets. However, for the machine learning models to be reliable, quality datasets representative of the target population need to be ensured~\cite{yu2018artificial}. 
The labels in these datasets are often generated from human annotations using automated extraction or entity detection tools. 
However, these annotations (and their structured archiving) from automated tools can have errors due to faulty perceptions, interpretations, and human errors~\cite{waite2017interpretive}. 
Even if the error rate of the annotator is less than 4\%, this can lead to millions of annotation errors per year~\cite{bruno2015understanding}. 
This parallels the 3.3\% error rate in large computer vision dataset~\cite{northcutt2021pervasive}.
Thus, there needs to be a better way to identify such errors before they are included in a dataset.

\begin{figure*}[t]
\centering
\includegraphics[width=0.5\linewidth]{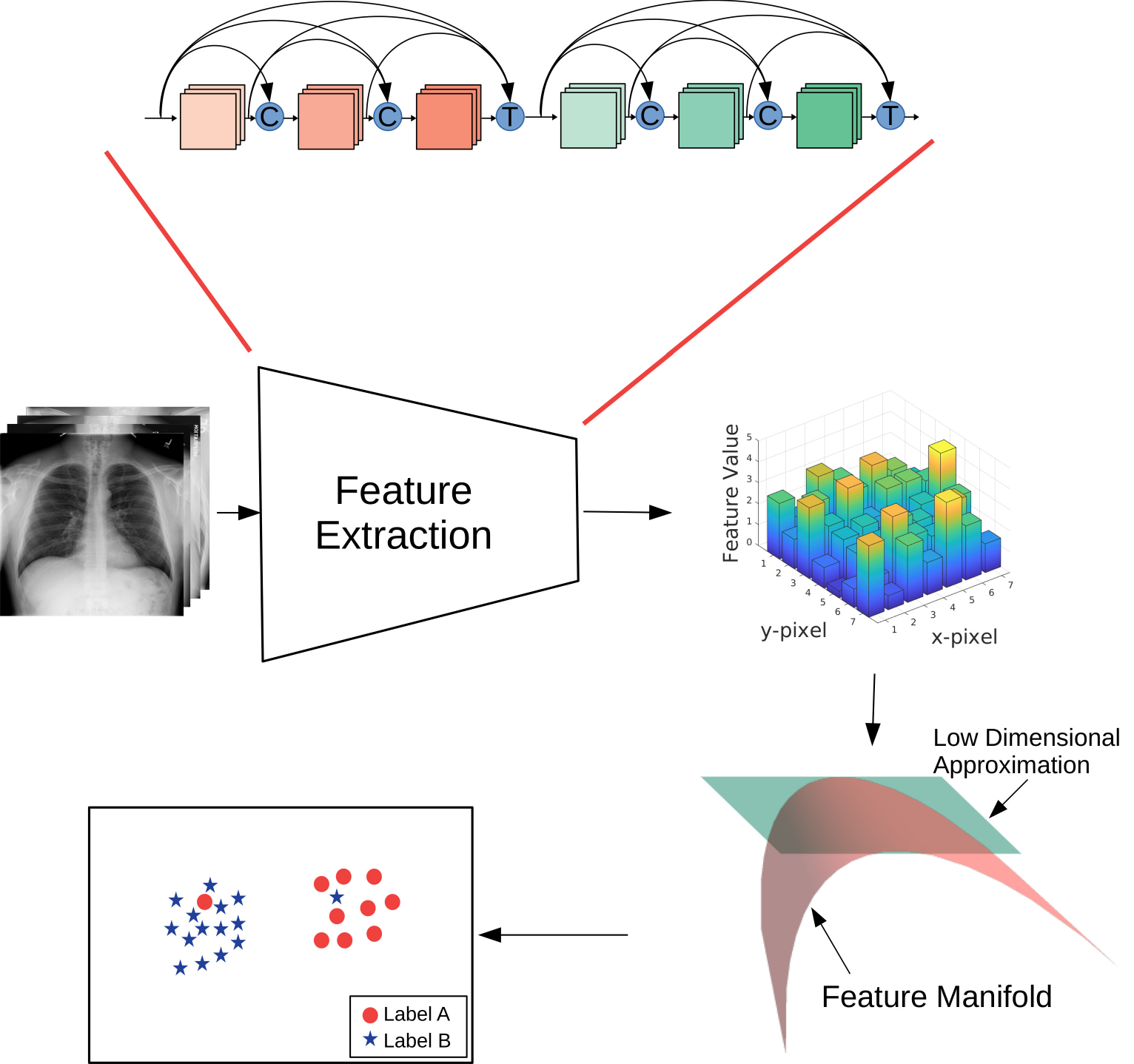}
\caption{Schematic of the outlier search algorithm. Image features extracted from a DenseNet-121 neural network are projected onto a low-dimensional space (2-D plane) using UMAP.}\label{fig:schematic}
\end{figure*}

For images, the search can be performed visually. However, examining individual images is a daunting task that requires many human hours. 
A popular alternative is neighbor embedding~\cite{hinton2002stochastic}, which can produce a two-dimensional (2-D) cluster plot that can be analyzed visually quickly. 
(This class of methods is also known as nonlinear dimensionality reduction as the 2-D plot preserves the pairwise similarity, i.e., graph structure, of the original high-dimensional space.)
Widely used neighbor embedding algorithms are t-distributed stochastic neighbor embedding (t-SNE)~\cite{maaten2008visualizing} and uniform manifold approximation and projection~(UMAP)~\cite{mcinnes2018umap}. 
UMAP was introduced relatively recently and has become very popular, as this method draws concepts from rich algebraic and topological structures and is computationally fast.

In this paper, we design a UMAP-based visual analytic method for extracting outlier images from large x-ray datasets. 
We validate our method by analyzing three publicly available and widely used medical image datasets. 
We show that the method can successfully cluster image features and produce interpretable visualization. 
We also discover labeling errors and erroneous images that have slipped through the verification process done prior to dissemination. Codes to reproduce the results are available at \url{https://github.com/tariqul-islam/Outlier_Detection_UMAP}.

\section{Related Works}\label{sec:relatedworks}

In the literature, the term outlier is often used interchangeably with abnormality and anomaly~\cite{fritsch2012detecting}.
Here, we define outliers as images that do not have sufficient signal for final decision-making or do not belong in the dataset due to specification. 
Generally, outlier detection methods assume an underlying distribution and often model it as normal distribution~\cite{hodge2004survey,han2022adbench}, i.e., a data point is an outlier if it is far away from the mean of the fitted distribution.
Fritsch et al.~\cite{fritsch2012detecting} used the minimum covariance determinant estimator and its extensions to find outliers (due to motion or registration issues) in neuroimaging data by analyzing principal components.
Gang et al.~\cite{gang2018dimensionality} used a t-SNE plot to find outliers from binary lung masks in terms of size variation and segmentation error. Fleischer and Islam~\cite{fleischer2020late} employed UMAP on chest x-rays for phenotyping COVID-19 response.

\section{System Overview}\label{sec:system_overview}

Following preprocessing (discussed in \ref{appendix:methods_outlier_2}), the major parts of the framework are feature extraction and dimensionality reduction (Fig.~\ref{fig:schematic}). To extract features from these images, we employed DenseNet-121~\cite{huang2017densely} trained on ImageNet~\cite{russakovsky2015imagenet}, a widely used deep neural network architecture designed to efficiently propagate features from earlier layers of a network to deeper layers. 
Importantly, neural algorithms are usually robust to many variabilities in images by design, and thus can accommodate standard images and outliers on equal terms.

Medical images usually vary in resolution, have different contrast, brightness, and alignment, and often suffer from registration issues.
In our framework, the features have been extracted from the final layer (before the softmax layer) of the network, where the features are generally most discriminating. 
Since we are not using a radiologically pre-trained model, these features generally will not be able to identify individual diseases. Rather, we employ other related labels (e.g., x-ray views, study labels) to examine the datasets. After extracting the features, we employ UMAP~\cite{mcinnes2018umap}, to obtain a 2-D approximation of the high-dimensional features.  

\section{Datasets}\label{sec:datasets}

We evaluate our approach on three publicly available datasets: ChestX-ray14~\cite{wang2017chestx}, CheXpert~\cite{irvin2019CheXpert}, and Musculoskeletal Radiographs (MURA)~\cite{rajpurkar2017mura}. 
ChestX-ray14 contains 112,120 frontal chest x-ray images from 30,805 unique patients. Images are from posterior-anterior (PA) and anterior-posterior (AP) views.
CheXpert dataset contains 224,316 chest x-rays (PA, AP, and Lateral) from 65,240 patients. We used 223,414 JPEG formatted x-rays from the training set of the dataset. 
MURA dataset contains 40,561 musculoskeletal x-rays from 14,863 studies. 
Like CheXpert, we used 36,808 x-rays from the training set and utilized study labels, e.g., finger, wrist, hand, forearm, elbow, humerus, and shoulder, for analysis.

\section{Experiments}\label{sec:experiments}

In this section, we start by analyzing embeddings of chest x-rays (ChestX-ray14 and CheXpert datasets) and find different types of outliers. 
Then, we expand on variations of our method by altering the neural architecture, pre-training dataset, and embedding algorithms. Finally, we discuss outliers in musculoskeletal x-rays of the MURA dataset.

\begin{figure*}[t]
\centering
\includegraphics[width=0.8\linewidth]{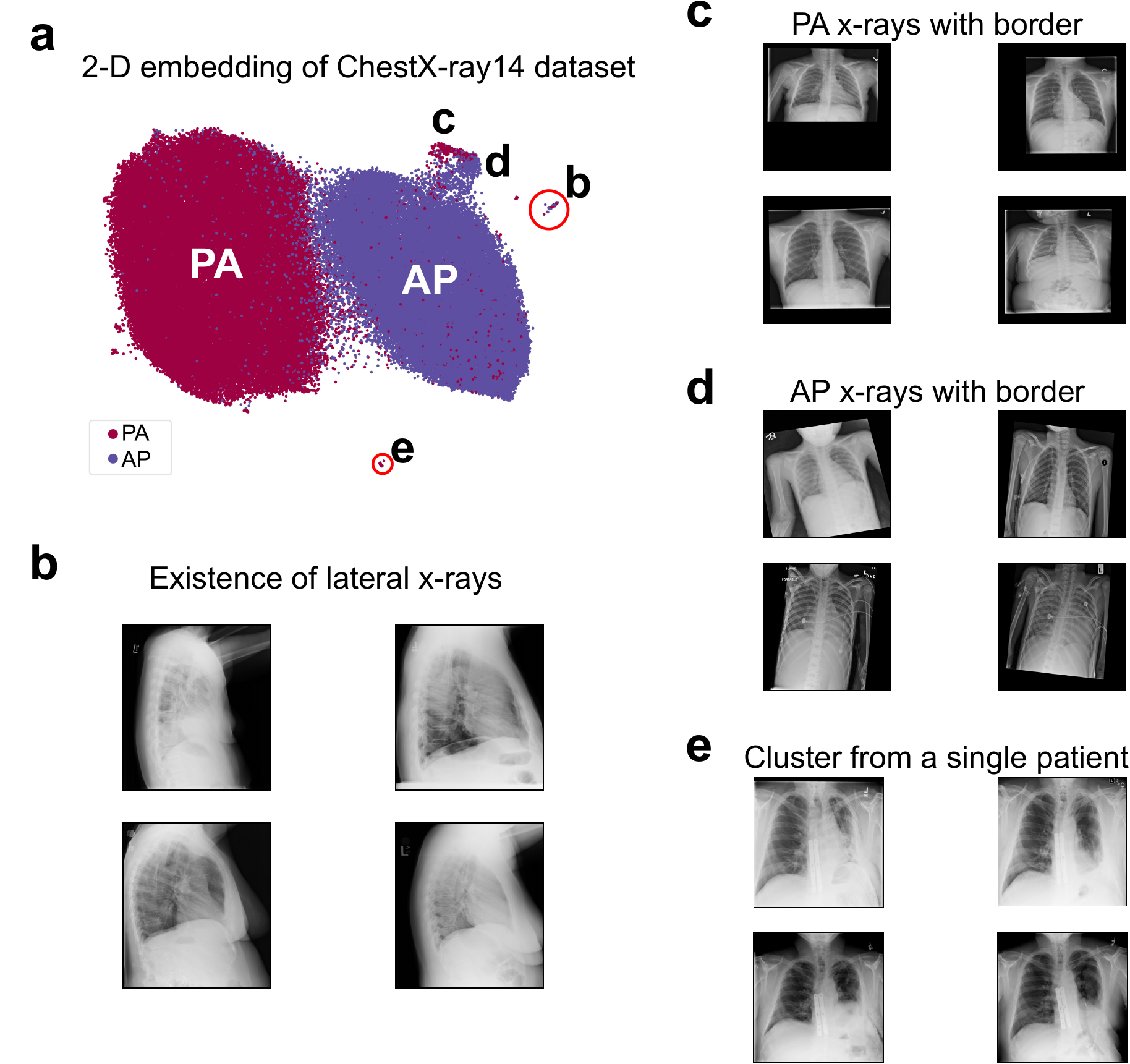}
\caption{Outlier detection in the ChestX-ray14 dataset. (a) 2-D embedding. Labeled clusters from (a) are: (b) Lateral x-rays which were not supposed to be in the dataset, (c) PA x-rays with borders, (d) AP x-rays with borders, and (e) cluster from a single patient.}\label{fig:cxr14}
\end{figure*}

\subsection{Analyzing Chest X-ray Datasets}

\subsubsection{Lateral X-rays in ChestX-ray14:}

\begin{figure*}[t]
\centering
\includegraphics[width=0.8\linewidth]{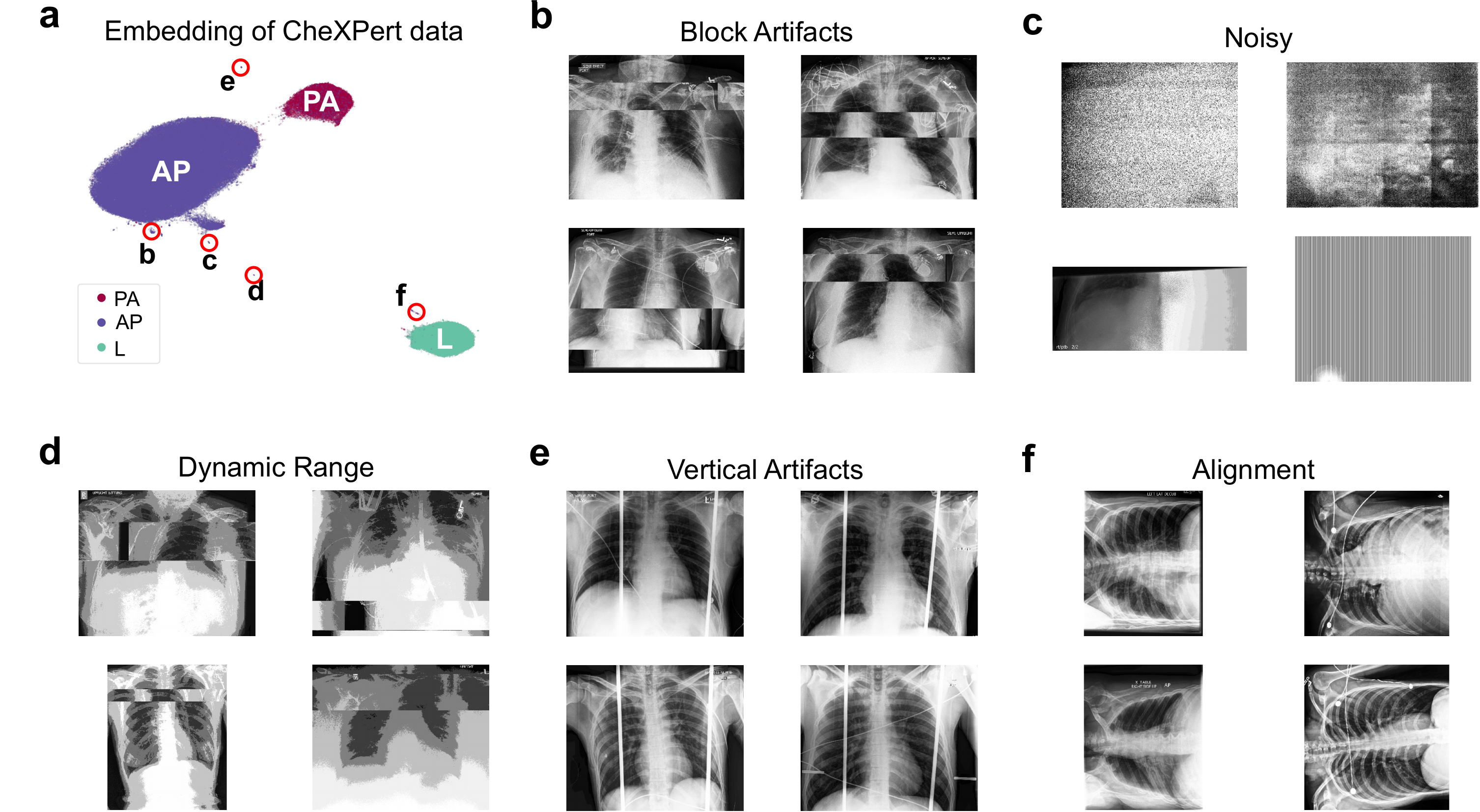}
\caption{Outlier detection in the CheXpert dataset. (a) 2-D Embedding. Example images with (b) block artifacts, (c) noise, (d) improper dynamic range, (e) vertical artifacts, and (f) alignment issues.}\label{fig:CheXpert_complete}
\end{figure*}

Our 2-D embedding (Fig.~\ref{fig:cxr14}~(a)) shows two large clusters of PA and AP views, corresponding to the primary topology of the DenseNet features. 
The embedding also includes a few satellite clusters around these large ones. 
These satellite clusters around the larger ones occur because the nearest neighbor graph creates a loop (or isolated sub-graph) of common features that are distinct from the rest of the data. 
In most cases, each of the satellite clusters of x-rays is from a single patient with a unique signature.
However, if any specific image features (such as similar artifacts in multiple images) are present in x-rays of different patients, these can create satellite clusters as well. Another interesting structure in Fig.~\ref{fig:cxr14}~(a) is the protruding region from the AP cluster.

Representative examples of anomalous clusters are shown in Figs.~\ref{fig:cxr14}~(b)-(e). 
The most surprising finding is the existence of some lateral x-rays in the dataset (Fig.~\ref{fig:cxr14}~(b)), as this dataset is supposed to be composed of frontal chest x-rays only. 
We found 92 lateral x-rays using our method. 

The protruding region from the AP cluster marked c and d 
(in Fig.~\ref{fig:cxr14}~(a)), consisting of x-rays with dark borders of PA and AP views, respectively. Finally, cluster (e) shown in Fig.~\ref{fig:cxr14}~(e) groups 46 x-rays from patient ID 9845, and a single x-ray from patient ID 12562.

\subsubsection{Corrupted Images in CheXpert:}

Figure~\ref{fig:CheXpert_complete}~(a) shows the 2-D embedding of CheXpert dataset. 
As before, the large PA and AP clusters form the bulk of the mapping. The lateral x-rays also form a separate large cluster. 
A few of the large satellite clusters (b-f) have been marked by red circles in Fig.~\ref{fig:CheXpert_complete}~(a). Four images from each of the clusters are plotted in Fig.~\ref{fig:CheXpert_complete}~(b)-(e). 
Figure~\ref{fig:CheXpert_complete}~(b) depicts images with block artifacts, e.g., from poor JPEG compression or accidental splicing. 
We found 107 such images in this cluster. 
Figure~\ref{fig:CheXpert_complete}~(c) depicts images that are just noise (19 images). 
Figure~\ref{fig:CheXpert_complete}~(d) shows images with block artifacts and dynamic range issues (53 images). 
Figure~\ref{fig:CheXpert_complete}~(e) shows x-rays with vertical artifacts (88 images).
Finally, Fig.~\ref{fig:CheXpert_complete}~(f) shows rotated images. 
This cluster is placed near the large cluster of lateral (L) x-rays. 
Thus, DenseNet considers rotated x-rays to be more similar to lateral images than upright frontal x-rays.

\begin{figure*}[th]
\centering
\includegraphics[width=1.0\linewidth]{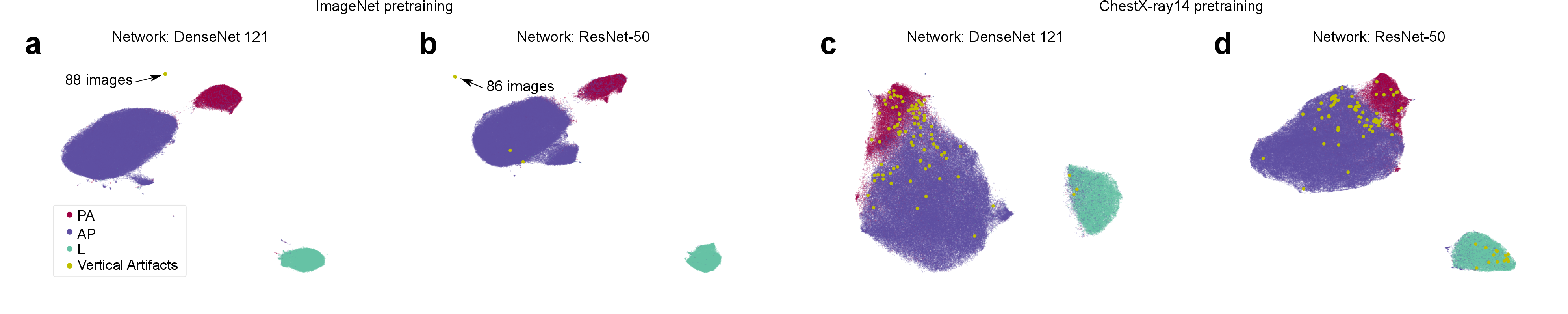}
\caption{Embedding of CheXpert dataset using different pre-trained models. DenseNet-121 and ResNet-50 trained on ImageNet (left two) and ChestX-ray14 (right tow) datasets. Each yellow point represents an image with vertical artifact (from cluster e in Fig.~\ref{fig:CheXpert_complete}~(a)) indicating chest x-ray pre-trained models fail to identify these as outliers.} \label{fig:different_networks}
\vspace{0.1in}
\centering
\includegraphics[width=0.5\linewidth]{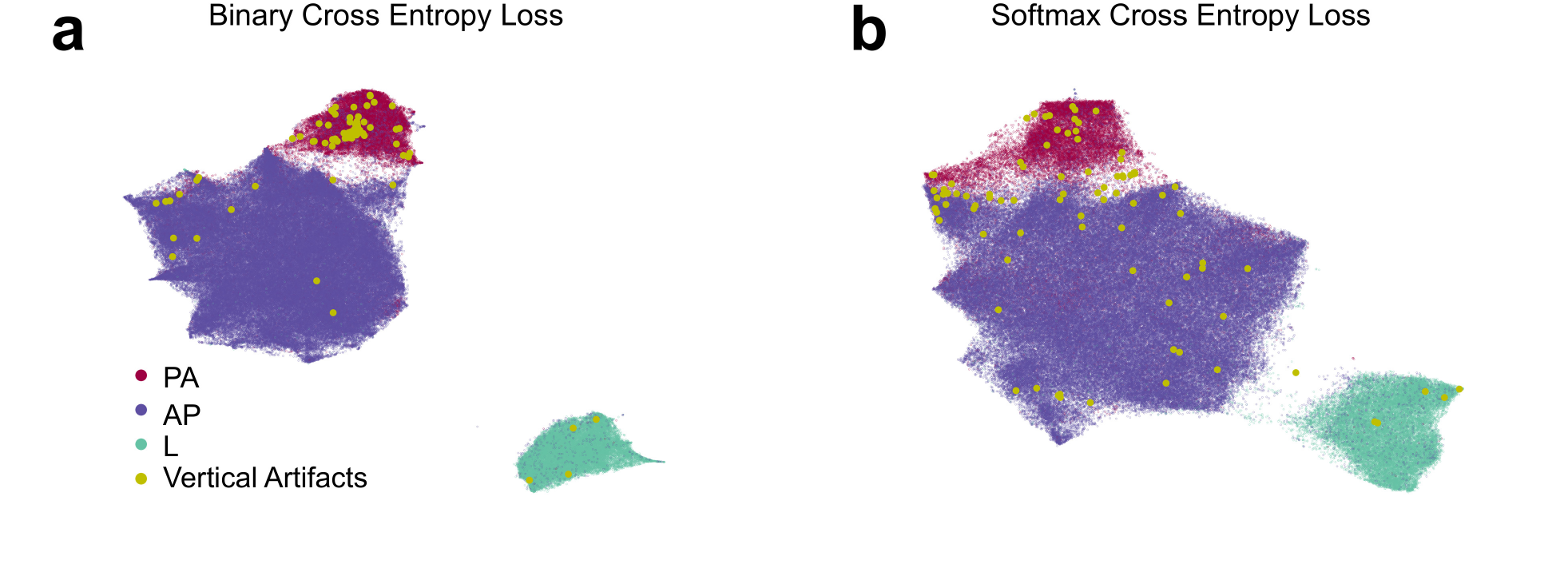}
\caption{Pre-training DenseNet-121 models using non-overlapping labels from ChestX-ray14 dataset and embedding of CheXpert dataset using UMAP. Models are trained using (a) binary cross entropy loss and (b) softmax cross entropy loss. Each yellow point represents an image with vertical artifacts (from cluster e in Fig.~\ref{fig:CheXpert_complete}~(a)).} \label{fig:different_networks_unique}
\end{figure*}

\begin{figure*}[t]
\centering
\includegraphics[width=1.0\linewidth]{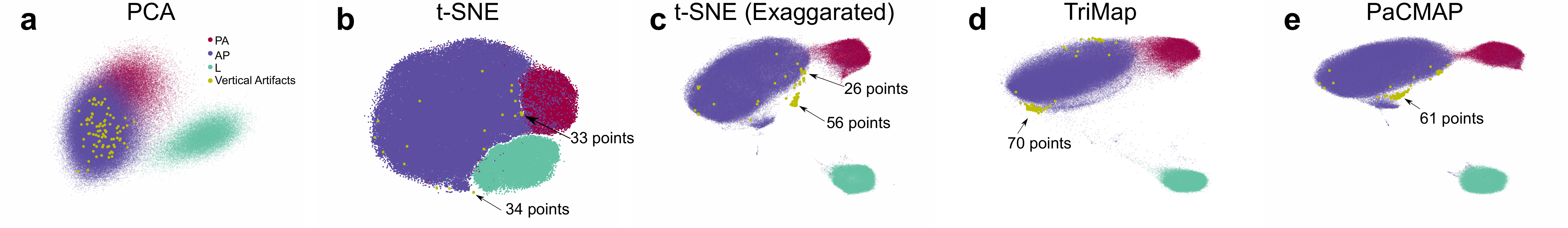}
\caption{Embedding of CheXpert dataset using several dimensionality reduction algorithms. (a) PCA, (b) t-SNE, (c) t-SNE (exaggerated), (d) TriMap, and (e) PaCMAP. Each yellow point represents an image with vertical artifacts (from cluster e in Fig.~\ref{fig:CheXpert_complete}~(a)).} \label{fig:different_embedders}
\end{figure*}

\subsection{Effect of Different Pre-trained Networks}

In the previous section, we used DenseNet-121 trained on ImageNet data for feature extraction (Fig.~\ref{fig:schematic}). 
However, using ImageNet may introduce domain shift, since the dataset is not specific to medical images. 
To assess, we looked at embeddings from different models trained on ImageNet and chest x-rays (Fig.~\ref{fig:different_networks}). 

ImageNet models (Fig.~\ref{fig:different_networks}~(a,b)) create similar clusters of PA, AP, and L x-rays, whereas Chest-Xray14 models (Fig.~\ref{fig:different_networks}~(c,d)) merge the PA and AP clusters. 
The x-ray model embeddings also contain fewer satellite clusters, and their distinctness is lost. Thus, outlier detection fails. 
Additionally, we highlight the placement of images with vertical artifacts from the cluster (e) of Fig.~\ref{fig:CheXpert_complete} in the alternate models (yellow points in Fig.~\ref{fig:different_networks}). 
While both ImageNet models successfully separate these images (with a few scattered around), all the chest x-ray models fail (where the outliers are inside the large cluster and scattered throughout). 

A major difference between models is that ImageNet models use softmax (n-ary) cross entropy loss (for non-overlapping labels), but chest-x-ray models use binary cross entropy loss (for overlapping labels). 
To resolve this discrepancy, we trained chest x-ray models using non-overlapping labels of the ChestX-ray-14 dataset (for details, see Appendix~\ref{appendix:pretrain_cxrmodel}). 
The resulting embeddings (Fig.~\ref{fig:different_networks_unique}) show a similar characteristic of the chest x-ray models with overlapping labels (Fig.~\ref{fig:different_networks}~(c,d)), in that the individual views are weakly separated while the specific outlier images fail to form the satellite clusters. 
For example, the outliers with vertical artifacts scatter within the 2D mapping. The networks consider them as any other x-ray image and ignore the artifacts. 
This result strengthens the idea that training on ImageNet (or a more general computer vision task) that has broader exposure benefits the discovery of outlier images. 

\subsection{Comparing Various Embedding Algorithms}

To assess the effectiveness of UMAP, we compared it with a few other dimensionality reduction techniques, specifically principal component analysis (PCA), t-SNE, TriMap~\cite{amid2019trimap}, and PaCMAP~\cite{wang2021understanding}.

The PCA embedding in Fig.~\ref{fig:different_embedders}~(a) shows the top two directions of largest variances. There are two clusters but the AP and PA views overlap. The nonlinear dimensionality reduction algorithms - t-SNE, UMAP, and variants - discover more features and make the clusters more distinct. 

The default t-SNE is tuned to preserve the neighborhood as best as possible. This often causes the individual clusters to spread out and be less compact (Fig.~\ref{fig:different_embedders}~(b)). This behavior is apparent in the PA, AP, and lateral x-ray clusters: the separation among them is minimal, and there is little room for the satellite clusters. 
However, t-SNE can be tuned to produce a more UMAP-like output. 
Following the findings of Linderman et al.~\cite{linderman2019clustering} and Bohm et al.~\cite{bohm2020unifying}, we used an exaggeration factor of 4, which means we applied four times more attractive force than the repulsive force throughout the optimization procedure. 
The standard early exaggeration factor of 12 was also applied at the start of the optimization. The resulting plot in Fig.~\ref{fig:different_embedders}~(c) largely resolves the PA and AP clusters, but there are fewer satellite clusters than the UMAP output.

Both PaCMAP~\cite{wang2021understanding} and TriMap~\cite{amid2019trimap} aim to preserve the global structure of the data while keeping the clustering properties of UMAP. PaCMAP modifies the pairwise relation of UMAP by considering different neighborhoods at different scales (near and far), while TriMap achieves this with the triplet constraint. These methods obtain the clustering of PA, AP, and lateral views (Figs.~\ref{fig:different_embedders}~(d,e)), but, similar to exaggerated t-SNE, the satellite clusters are largely absent.

The yellow dots show the placement of x-rays with vertical artifacts from cluster (e) of Fig.~\ref{fig:CheXpert_complete} for the different embedding algorithms. In all the alternate embedding algorithms, these images are scattered within the other x-ray images. The outlier images fails to be identified and labeled separately, which demonstrates a superior performance of UMAP.

\subsection{Extracting Mislabeled X-rays from MURA}\label{sec:seeding_data}

\begin{figure*}[t]
\centering
\includegraphics[width=0.8\linewidth]{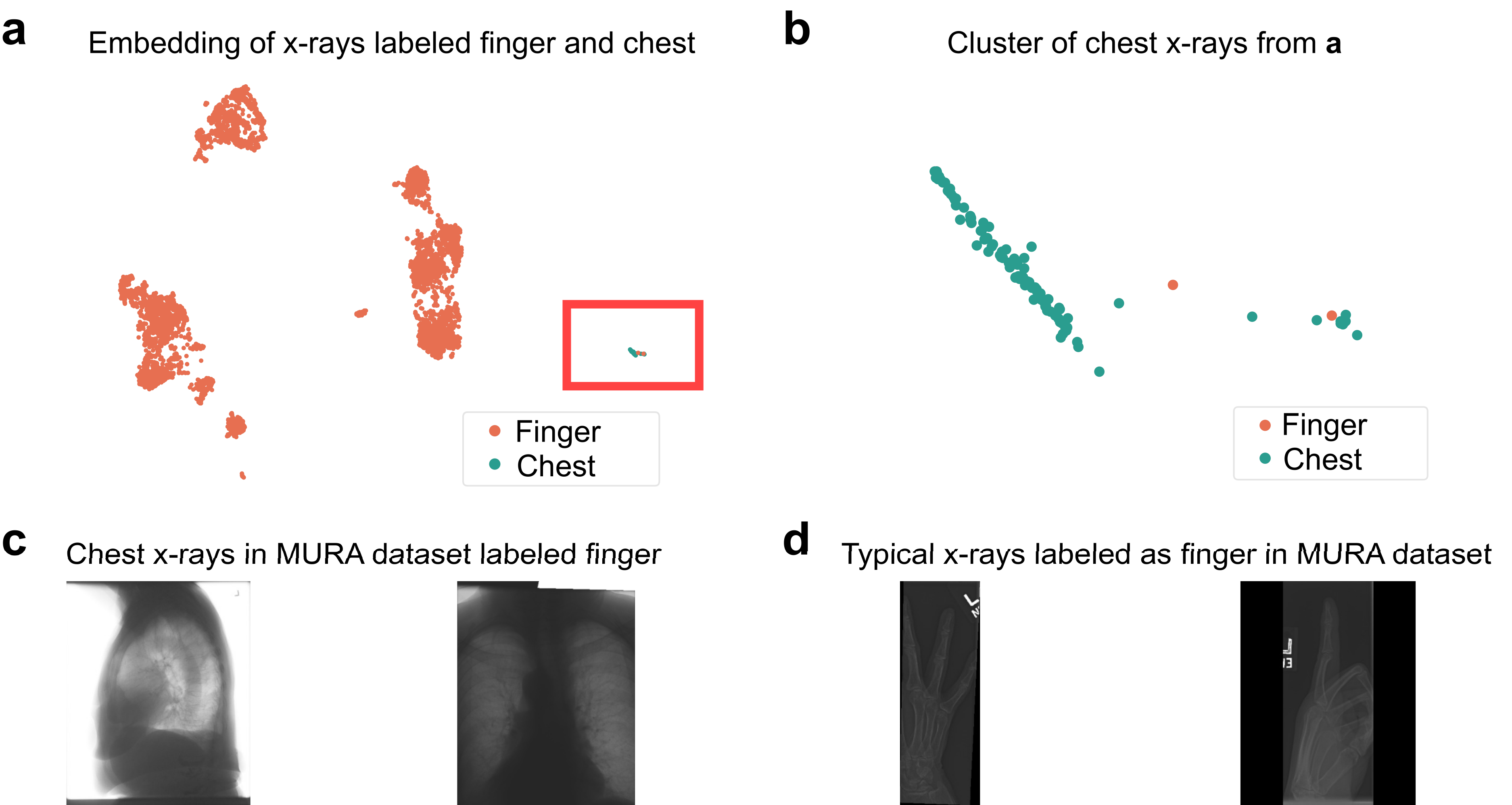}
\caption{Embedding of `finger' x-rays from MURA dataset and 100 chest x-rays from CheXpert dataset using UMAP. (a) Scatter plot of the embedding. The cluster of chest x-rays is marked using a red rectangle. (b) Scatter plot in the red rectangle. (c) two x-rays labeled `finger' are actually chest x-rays. (d) Typical finger x-rays from the MURA dataset.} \label{fig:MURA_finger}
\vspace{0.1in}
\end{figure*}

Since the MURA dataset consists of x-rays from different parts of the arm and the shoulder, there is a natural ambiguity in the labels, e.g., both wrist and hand x-rays may contain the hand of a person, and shoulder x-rays may contain part of the chest. 
In such cases, finding mislabeled x-rays by embedding all the images may be sub-optimal.
To find outliers more directly, we searched for misclassified images by explicitly using labels of the dataset. The method has two parts: 1) introduce target images with a specific label (preferably from a different dataset than the MURA one); and 2) perform neighbor embedding on the joint dataset.

For example, to look for possible chest x-rays that are falsely classified as finger x-rays, we added 100 chest x-rays from the CheXpert dataset to the 5,106 finger x-rays of MURA. 
We then applied the UMAP to the composite set (Fig.~\ref{fig:MURA_finger}~(a)). As shown in Figs.~\ref{fig:MURA_finger}~(b-d), the seeded chest x-rays acted as an attractor for mislabeled images in MURA, with x-rays labeled `finger' now appearing in the (new) chest cluster. 
Interestingly, both of these x-rays were from patient 04547 (another 3 from this patient were labeled correctly). We describe an additional experiment in Appendix~\ref{sec:seeding_data}.

\section{Conclusion}\label{sec:conclusion}

Neighbor embedding algorithms can be an effective tool for summarizing datasets and identifying outlier images. The principle of the method is that the outliers are different from the main data but can have similarities among themselves. 
Thus, the different outlier types form distinct clusters in the embeddings. 
Our experiments, using a DenseNet-121 feature extractor and UMAP neighbor embedding method on the ChestX-ray14, CheXpert, and MURA datasets, distinguished different radiological views of chest x-rays, classified differences, and identified wrongly labeled or corrupted images.  We further found specific types of outliers by seeding the dataset with target images and performing neighbor embedding.

While this study performed retrospective analysis of large x-ray datasets, outlier curation can be achieved during the initial assembly of the dataset as well. 
For suspected outliers, the method of seeding data with known labels can be applied. To streamline the process, appropriate reference datasets may be created beforehand. Undoubtedly, cleaner input data will result in cleaner output data. For larger datasets, more accurate results and faster embedding may be achieved by dividing them into smaller subsets and applying better alignment techniques~\cite{islam2022manifold}. Finally, since the methods are graph-based and agnostic to the underlying data type, all of the methods here can be applied to arbitrary datasets, including and especially those that are mixed modality.

\subsubsection*{Acknowledgement} The authors gratefully acknowledge financial support from the Schmidt DataX Fund at Princeton University made possible through a major gift from the Schmidt Futures Foundation.

\subsubsection*{Disclosure}
The authors have no competing interests to declare that are relevant to the content of this article.


%
%
%
 \bibliographystyle{unsrt}
 \bibliography{thesis}

\appendix
\section{Implementation Details}\label{appendix:methods_outlier_2}

\subsection{Image Pre-processing}
We apply the following transformations: histogram equalization~\cite{e2008digital}, resizing the images, center cropping, and normalization. Images are resized such that the lowest dimension contains 256 pixels and then center-cropped to a $224\times224$ dimensional image for feature extraction. 
Then the images are normalized according to the specification of ImageNet: mean (0.485, 0.456, 0.406) and standard deviation (0.229, 0.224, 0.225) of red, green, and blue channels, respectively.

\subsection{Feature Extraction}
We use a DenseNet-121 architecture~\cite{huang2017densely} pre-trained on the ImageNet dataset~\cite{russakovsky2015imagenet} from the PyTorch deep learning library~\cite{paszke2017automatic}. (DenseNet is a deep neural network with many inter-layer connections designed to reduce the numerical instabilities that originate due to the depth of the network. The usage of neural network outputs as features is an effective baseline in machine learning algorithms~\cite{sharif2014cnn} and is extensively used in medical image analysis~\cite{cohen2022torchxrayvision}.)  Here, we remove the classification (softmax) layer from the neural network and use the output as features. 

\subsection{Parameter Settings of Embeddings}

\begin{figure*}[t]
\centering
\includegraphics[width=0.5\linewidth]{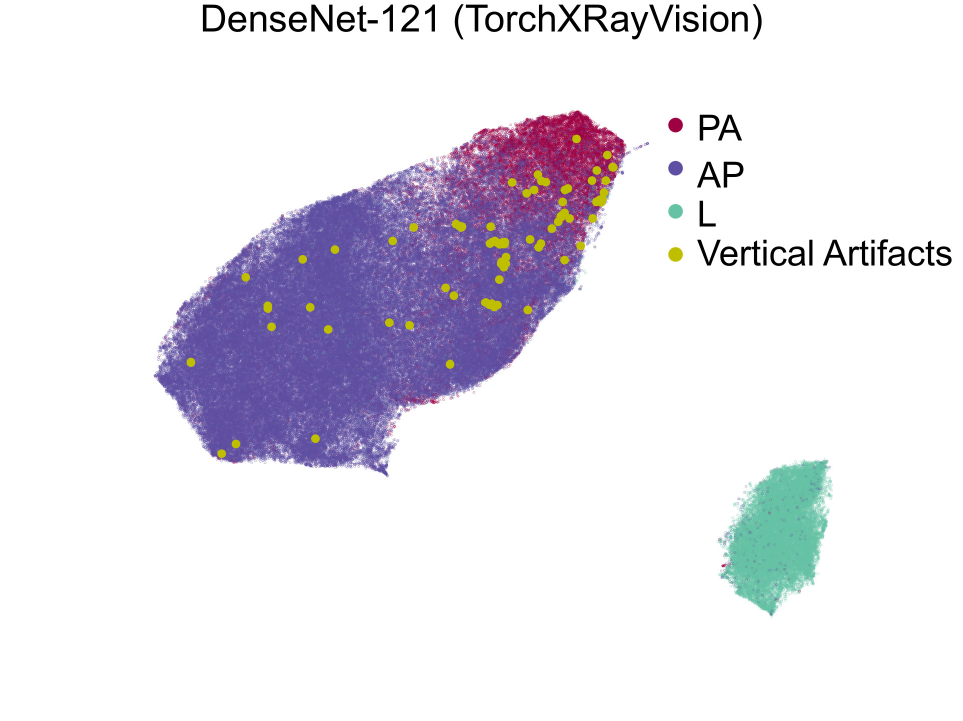}
\caption{Embedding of CheXpert dataset using publicly available DenseNet-121 model pre-trained on ChestX-ray14 dataset (from TorchXRayVision library~\cite{cohen2022torchxrayvision})} \label{fig:comparison_txr}
\end{figure*}

In general, we kept the number of nearest neighbors $k$ to be low, as increasing $k$ increases the computational budget exponentially.

ChestX-ray14 - Fig.~\ref{fig:cxr14}: For ChestX-ray14 we used $k=50$. The minimum distance parameter $m_d$ was set to $0.1$. We experimented with smaller $m_d$ values to increase the separation of the AP and PA x-rays, but it had little effect. The embedding was optimized for 200 epochs.

CheXpert - Figs.~\ref{fig:CheXpert_complete},~\ref{fig:different_networks}, and~\ref{fig:different_networks_unique}: The embedding was obtained by using $k=10$. We used a smaller minimum distance $m_d=0.001$, as we found that this value provided a better separability of the large clusters. We ran the optimization for 300 epochs.
Figure~\ref{fig:different_embedders}: (b) t-SNE: early exaggerated for 250 steps, then 500 regular steps, (c) t-SNE: early exaggerated for 250 steps, then 500 exaggerated steps, (d) TriMap: 1000 steps with $k=10$, compactness parameter $=0.1$, and (e) PaCMAP: 450 steps with $k=10$.

MURA - Figs.~\ref{fig:MURA_finger}, \ref{fig:MURA_embedding}, and \ref{fig:MURA_pair_process}: Similar to CheXpert, we used $k=10$, $m_d=0.001$, and ran the optimization for 300 epochs.

\section{Pre-training Models using ChestX-ray14 Dataset}\label{appendix:pretrain_cxrmodel}

We trained several chest x-ray models using the ChestX-ray14 dataset. We trained traditional x-ray models using binary cross entropy loss for every disease (one vs all model) by following~\cite{rajpurkar2017chexnet} which matches with the settings of the ImageNet pre-training. Additionally, to test whether the loss function and multilabel training have major effects on the resulting embeddings, we constructed a dataset by removing overlapping labels (Table~\ref{aptable:cxr14labels}) and trained two models using binary and softmax cross-entropy loss, respectively. For each mode, we optimized for 400,000 iterations using Adam (learning rate $=0.001$) with a batch size of 30 (2 images per disease label).

\begin{table*}[t]
 \caption{ChestX-ray14 labels and samples.} \label{aptable:cxr14labels}
  \centering
  \begin{tabular}{l|l|l}
    \toprule
    Label  & No. of labels & No. of non-overlapping labels  \\
    \midrule
    Atelectasis         & 11559     & 4215      \\
    Cardiomegaly        & 2776      & 1093      \\
    Effusion            & 13317     & 3955      \\
    Infiltration        & 19894     & 9547      \\
    Mass                & 5782      & 2139      \\
    Nodule              & 6331      & 2705      \\
    Pneumonia           & 1431      & 322       \\
    Pneumothorax        & 5302      & 2194      \\
    Consolidation       & 4667      & 1310      \\
    Edema               & 2303      & 628       \\
    Emphysema           & 2516      & 892       \\
    Fibrosis            & 1686      & 727       \\
    Pleural Thickening  & 3385      & 1126      \\ 
    Hernia              & 227       & 110       \\
    No Finding          & 60361     & 60361     \\
    \midrule
    Total Images        & 112,120   & 91324     \\
    \bottomrule
  \end{tabular}
\end{table*}

\subsection{Comparison to Publicly Available Chest X-ray Models}\label{appendix:publiccxr}

We compare our DenseNet-121 model to that from TorchXRayVision library~\cite{cohen2022torchxrayvision}. It differs from ours in that the image pixels are normalized to $[-1024,1024]$, images are converted to (or read as) grayscale, and `No Finding' label is untrained. 
This model (Fig.~\ref{fig:comparison_txr}) agrees with our results and ignores the artifacts of the outlier images, showing embeddings are indifferent to alternate preprocessing scheme.

\section{Additional Discussion on MURA}

\begin{figure}[t]
\centering
\includegraphics[width=0.5\linewidth]{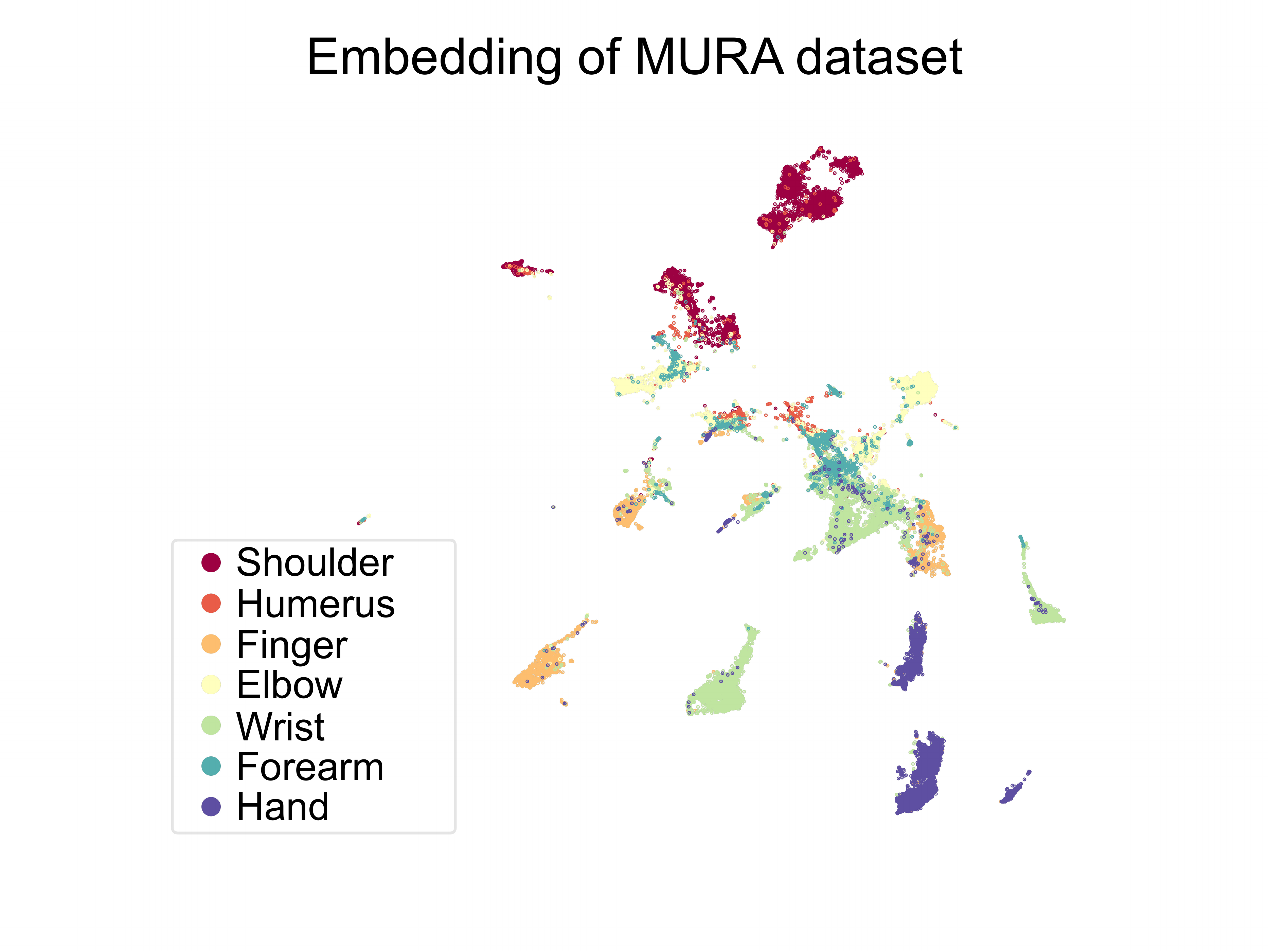}
\caption{Embedding of MURA dataset.} \label{fig:MURA_embedding}
\end{figure}

\begin{figure}[t]
\centering
\includegraphics[width=0.5\linewidth]{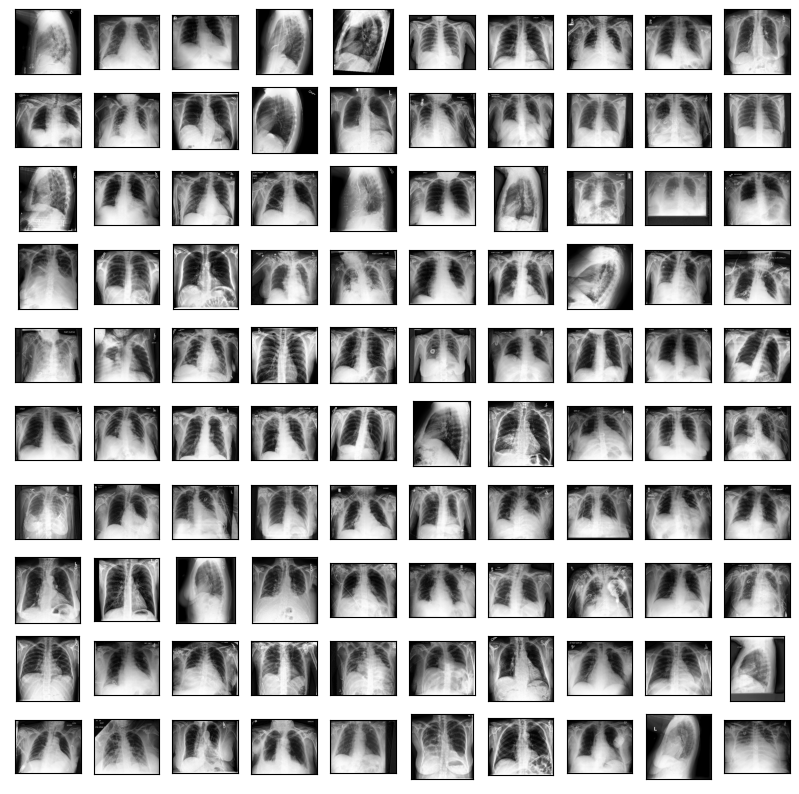}
\caption{Chest x-rays chosen from CheXpert dataset to produce the embedding in Fig~\ref{fig:MURA_finger}.} \label{fig:MURA_chestxrays}
\end{figure}

Fig.~\ref{fig:MURA_embedding} shows the embedding of 36,808 musculoskeletal radiographs from the MURA dataset. There is decent separation among the x-rays in terms of the labels, but there is also a considerable overlap. For example, finger, wrist, and hand x-rays overlap, as finger x-rays include parts of the wrist and hand, and vice versa. Similarly, wrist and forearm clusters are often merged, since x-ray of the forearm tend to capture a portion of the wrist, and vice versa. The same happens for humerus and shoulder. The images within each cluster thus share similar acquisition views, aspect ratios, and specific features (e.g., circular window function, stitching of multiple x-rays in one image). 
Unlike ChexPert case, analysis of this mapping did not reveal any satellite clusters with corrupted images.

Based on this, we decided to focus on individual labels and extract images with specific labels. The chest x-rays from the CheXpert dataset used to produce Fig.~\ref{fig:MURA_finger} are shown in Fig.~\ref{fig:MURA_chestxrays}.

\section{Extracting Mislabeled X-rays from MURA}\label{sec:seeding_data}

\begin{figure*}
\centering
\includegraphics[width=0.8\linewidth]{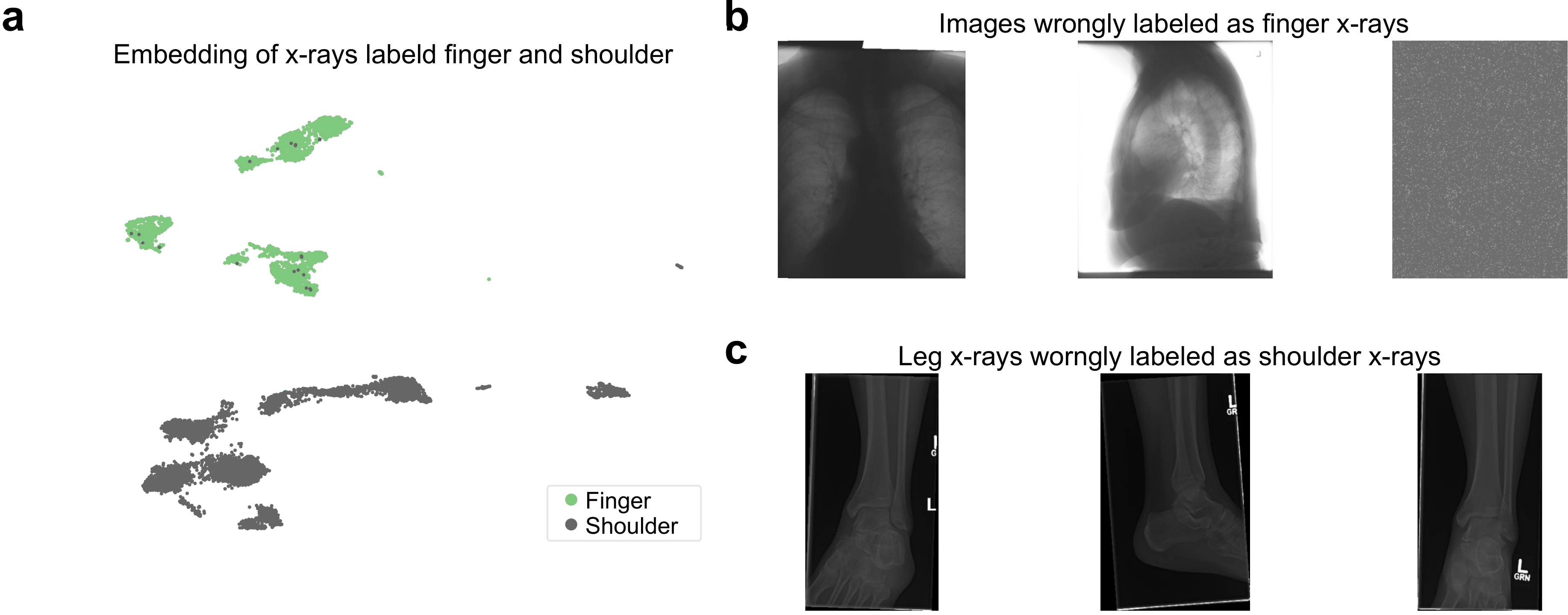}
\caption{Embedding of `finger' and `shoulder' x-rays from MURA dataset using UMAP. (a) 2-D scatterplot of the embedding. (b) Chest x-ray and non-x-ray images were discovered which are labeled as `finger' x-rays. (c) Leg x-rays labeled as `shoulder' x-rays.}\label{fig:MURA_pair_process}
\end{figure*}

In a different experiment on the MURA dataset, we used `finger' and `shoulder' x-rays (Fig.~\ref{fig:MURA_pair_process}). 
The broad features of the finger and shoulder are easily separable with a few misclassified points. 
Analyzing `finger' x-rays misclassified in `shoulder' clusters, we can find the two chest x-rays labeled as finger (which we found in previous experiment as well) and two images that are just noise/non-x-ray images (Fig.~\ref{fig:MURA_pair_process}~(b)). 
The latter belongs to patient ID 04687, which uncovers two more outliers (one is an x-ray of keys). The misclassified `shoulder' labels in the `finger' cluster reveal three leg x-rays (Fig.~\ref{fig:MURA_pair_process}~(c)).

\end{document}